\documentclass[aps,article,twocolumn,superscriptaddress,english]{revtex4}

\usepackage{graphicx}
\usepackage{dcolumn}
\usepackage{bm}
\graphicspath{ {./images/} } 
\usepackage{subcaption}
\usepackage{caption}
\usepackage{comment}
\usepackage{multirow}
\usepackage{color}
\usepackage{graphics,epsfig,amsfonts,amssymb,amsmath,color}

\begin{document}

\title{Magnetic field regression using artificial neural networks for cold atom experiments}

\author{Ziting Chen}
\thanks{These authors contributed equally.}
\affiliation{Department of Physics, The Hong Kong University of Science and Technology, Clear Water Bay, Kowloon, Hong Kong, China}

\author{Kin To Wong}
\thanks{These authors contributed equally.}
\affiliation{Department of Physics, The Hong Kong University of Science and Technology, Clear Water Bay, Kowloon, Hong Kong, China}

\author{Bojeong Seo}
\affiliation{Department of Physics, The Hong Kong University of Science and Technology, Clear Water Bay, Kowloon, Hong Kong, China}

\author{Mingchen Huang}
\affiliation{Department of Physics, The Hong Kong University of Science and Technology, Clear Water Bay, Kowloon, Hong Kong, China}

\author{Mithilesh K. Parit}
\affiliation{Department of Physics, The Hong Kong University of Science and Technology, Clear Water Bay, Kowloon, Hong Kong, China}

\author{Haoting Zhen}
\affiliation{Department of Physics, The Hong Kong University of Science and Technology, Clear Water Bay, Kowloon, Hong Kong, China}

\author{Jensen Li}
\affiliation{Department of Physics, The Hong Kong University of Science and Technology, Clear Water Bay, Kowloon, Hong Kong, China}

\author{Gyu-Boong Jo}
\email{gbjo@ust.hk}
\affiliation{Department of Physics, The Hong Kong University of Science and Technology, Clear Water Bay, Kowloon, Hong Kong, China}

\begin{abstract}
Accurately measuring magnetic fields is essential for magnetic-field sensitive experiments in fields like atomic, molecular, and optical physics, condensed matter experiments, and other areas. However, since many experiments are conducted in an isolated vacuum environment that is inaccessible to experimentalists, it can be challenging to accurately determine the magnetic field. Here, we propose an efficient method for detecting magnetic fields with the assistance of an artificial neural network (NN). Instead of measuring the magnetic field directly at the desired location, we detect magnetic fields at several surrounding positions, and a trained NN can accurately predict the magnetic field at the target location. After training, we achieve a relative error of magnetic field magnitude (magnitude of error over the magnitude of magnetic field) below 0.3$\%$, and we successfully apply this method to our erbium quantum gas apparatus. This approach significantly simplifies the process of determining magnetic fields in isolated vacuum environments and can be applied to various research fields across a wide range of magnetic field magnitudes.
\end{abstract}

\maketitle


\section*{Introduction}

Precisely calibrating the magnetic field inside a vacuum chamber is of experimental significance to various fields. For instance, a magnetic field is one of the common control knob in the ultracold atoms experiment, enabling various studies on many-body physics~\cite{bloch2008many} including BEC-BCS crossover~\cite{bourdel2004experimental}, formation of matter-wave soliton~\cite{khaykovich2002formation,strecker2002formation,burger1999dark}, and Efimov states~\cite{kraemer2006evidence} through the Feshbach resonance that tunes inter-atomic interactions~\cite{chin2010feshbach}. However, the demanding precision of the magnetic field imposed by these experimental control and the inaccessibility to the vacuum chamber makes calibration of the magnetic field a difficult and time-consuming task. Moreover, spectroscopic measurement,  a typical approach to magnetic field calibration, is mainly sensitive only to the magnitude of the magnetic field. The magnetic field direction and its precise calibration are of critical importance for magnetic atoms (e.g. erbium or dysprosium) where the orientation of the magnetic dipole moment plays a critical role~\cite{lahaye2009physics,chomaz2022dipolar}.

Recent years have witnessed the great success of neural network (NN) applied to assist experiments, including multi-parameter optimization of magneto-optical trap~\cite{tranter2018multiparameter,seo2021maximized}, optimization of production of Bose-Einstein condensate~\cite{wigley2016fast,barker2020applying,vendeiro2022machine,davletov2020machine}, and recognition of hidden phase from experimental data~\cite{zhang2019machine,zhao2021heuristic,zhao2022observing,guo2021machine}. Here, we introduce a novel method to precisely determine the magnetic field vector $B=(B_x, B_y, B_z)$ inside a vacuum chamber with the assistance of a NN. Since the target position inside the vacuum chamber is typically inaccessible, we detect magnetic fields at several surrounding positions, which are sent to the trained NN that is able to accurately deduce the magnetic field inside the vacuum chamber. We apply this method to our apparatus of erbium quantum gas~\cite{seo2020efficient,seo2023apparatus,chen2021active}, which has large magnetic dipole moment making it particularly sensitive to magnetic field vector.~\cite{aikawa2012bose,frisch2014quantum}. We present the details of the NN-based method, including setting up the simulation model, training process, and final performance.  For simplicity, magnetic field data for training and validation of NN is generated by a standard finite-element simulation package from COMSOL Multiphysics electromagnetic module~\cite{pepper2017finite}, instead of experimental measurement. Moreover, we systematically investigate the impact of the number of sensors at surrounding positions and the magnitude of the magnetic field on the performance of the method providing a practical guide for implementation. In contrast to previous works~\cite{solin2018modeling,nouri2015prototype,nouri2014systematic,raissi2019physics,coskun2022magnetic}, which predict the magnetic field vector across a wide experimental region, our goal in this work is to extrapolate the magnetic field vector at a specific position within an inaccessible region. Our approach provides a simple method for monitoring magnetic fields without requiring any prior knowledge of solution of the Maxwell equation.

\begin{figure*}
\includegraphics[width=1\linewidth]{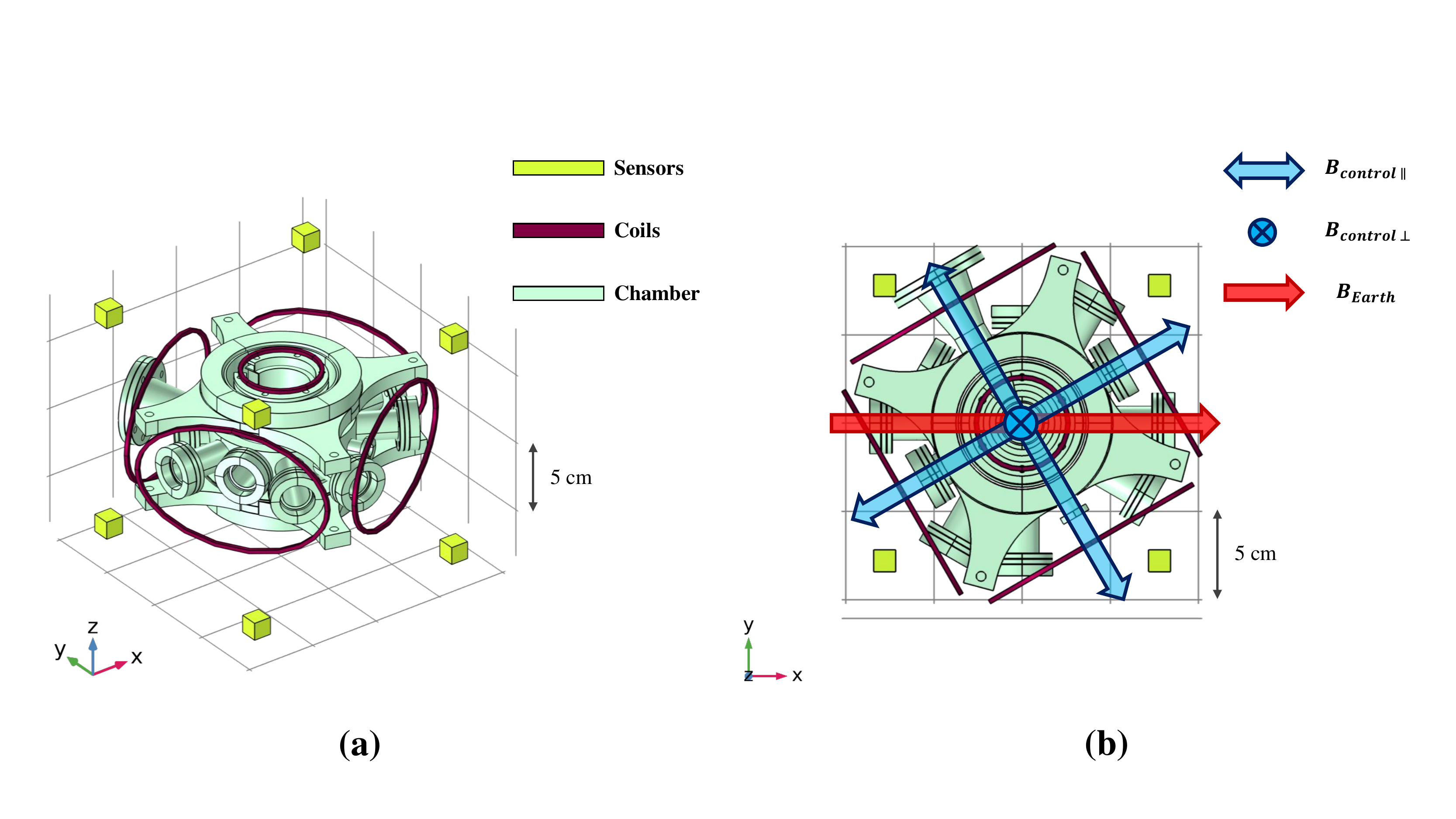}
\centering
\caption{ {\bf Schematic of the simulation model.} (a) Overview of the model. The main body of the model is an experimental chamber in ultra-high vacuum environment. The control magnetic field $B_{control}$ is generated by three orthogonal pairs of copper coils. Several magnetic field sensors surrounding vacuum chamber are indicated in light yellow. (b) Top view of the model. An exemplary magnetic fields (blue arrows) generated by coils are shown. A red arrow is also added to show the Earth's magnetic field in the model which is set to the x direction.}
\label{Fig1}
\end{figure*}

\section*{Methodology}

The implemented machine learning algorithm is an artificial NN that can be coded by common Python packages like Tensorflow and Pytorch \cite{tensorflow,pytorch}. The magnetic field measured by sensors outside a vacuum chamber is fed into the NN, which predicts the magnetic field at the center of the chamber. Hence, the function of the NN is to act as a hyperplane that relates the magnetic field at different spatial points. Regressing such a hyperplane would require a substantial amount of data. Obtaining training data from a real experimental setup, though it can take into account every factor, is difficult and time-consuming. Instead, we generate training data from finite element simulation using COMSOL Multiphysics. As long as the simulation model takes into account important factors, simulation data would be a reliable substitute for real data.

The simulation model is shown in Fig.~\ref{Fig1}, which contains a science chamber commonly used in cold atoms experiments. Originally, the chamber is made of materials including 314L, 304 stainless steel, aluminum, glass, and plastic. However, to reduce computational time, non-magnetic parts of the chamber are removed as they do not affect the result, and only parts made of 314L, and 304 stainless steel are kept. 314L and 304 stainless steel, under low strain and irradiation, can be considered as linear materials with no hysteresis loop~\cite{Xu2018}. In addition, only a static magnetic field is of concern for most experiments. Therefore, we only need to simulate the magnetic field under time-independent conditions. In order to generate a stable magnetic field, three pairs of copper coils with constant current are placed around the chamber along three orthogonal directions, mimicking the coils used in experiments. To account for the Earth magnetic field $B_{Earth}$ in the laboratory, a constant 400~mG magnetic field is added in the x-direction. To notice, when adopting this method, it is important to check the direction and magnitude of $B_{Earth}$ before training the NN since they can vary in different locations.

\begin{figure*}
    \includegraphics[width=1\linewidth]{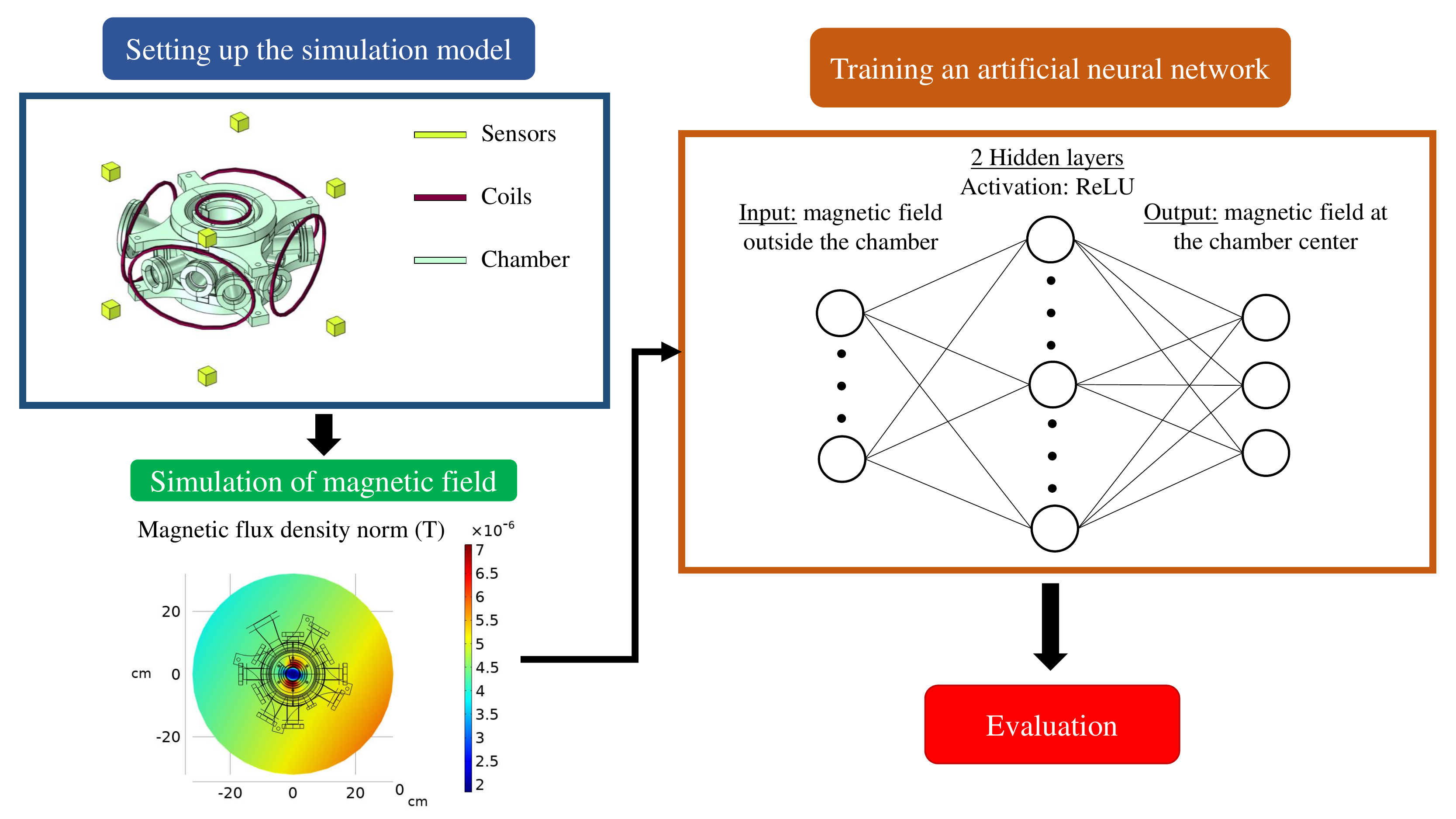} 
    \centering
    \caption{{\bf Pipeline figure for the overall procedure.} First, a simulation model which contains a science chamber and coils as magnetic field sources are set up. Then, the model is simulated to obtain data that is passed into an artificial NN for training. The NN has four layers: the input layer which is the magnetic field measured by sensors outside the chamber, two fully connected hidden layers with ReLU as activation functions, output layer which is the target magnetic field at the chamber center. At last, the prediction of the NN is evaluated.}
    \label{Fig2}
\end{figure*}

Even though simulation data is much easier to obtain than real data, simulation of each datum still requires a significant amount of time. To diminish data acquisition time, we exploit the linearity of Maxwell's equations that any superposition of solutions is also a valid solution for linear materials. By virtue of this, the whole output space can be mapped using only three linear independent results, which can be obtained by simulating each pair of coils in the model. Using these results, we obtain new data that is not redundant and form a large data set.

As shown in Fig.~\ref{Fig2}, after setting up the simulation model and obtaining the simulation data, we input the data into the artificial NN for training and prediction testing. The implemented NN contains two fully-connected hidden layers with ReLU as activation functions. During the training, the neurons' parameters of the NNs are adjusted to minimize the root-mean-square error (RMSE) loss function using the Adam optimizer~\cite{zhao2022observing}. This is a common loss function for regression and can be defined as $\mathbf{RMSE} = \sqrt{\frac{1}{3n}\sum_{i}^{n} \sum_{\alpha}^{x,y,z} (y^{\alpha}_{i}-\hat{y}^{\alpha}_{i})^{2}}$ where $n$ is the number of data, $\hat{y}$ is the actual output, $y$ is the predicted output and $\alpha$ denotes the component of the magnetic field data. Total number of simulation data is 1.5$\times$10$^5$, 80$\%$ of which is used for training while 20$\%$ is reserved for validation. When the validation loss isn't improving for several epochs, the training is terminated to prevent overtraining. It should be noted, however, RMSE could not reflect the statistic of the error for each measurement. 

To properly evaluate the performance of the NN-based magnetic field regression, we define the relative prediction error which is defined as the vector difference between the actual output and prediction over the magnitude of the actual output:
\begin{equation}
    \text{Relative prediction error} = \frac{||y-\hat{y}||}{||\hat{y}||}
\end{equation}
This value is evaluated from another data set of size of 10$^5$. We calculate the error from each datum to form a  set of errors and extract the upper bound of the relative error, below which 90$\%$ of data points would be accurately estimated.

\section*{Interpretation of the training process}
To fully evaluate the NN's performance, it is essential to first grasp the purpose of training the NN or specifically, the target equation of the regression. In theory, the hyperplane relating magnetic field at different spatial points can be calculated using Maxwell's equations directly. By substituting the spatial coordinates of the target point into the general solution, we remove its spatial degree of freedom, causing the output to depend only on the integration constants that are determined by the boundary conditions. However, the boundary conditions can just be the magnetic field at other spatial points. Hence, in principle, we can extract a hyperplane from the solution that serves the same function as the NN. Since this hyperplane is entirely based on Maxwell's equations, its output must always be correct. Therefore, the goal of training the NN is to reduce the gap between the NN and this ideal hyperplane.

\begin{figure}
    \includegraphics[width=0.7\linewidth]{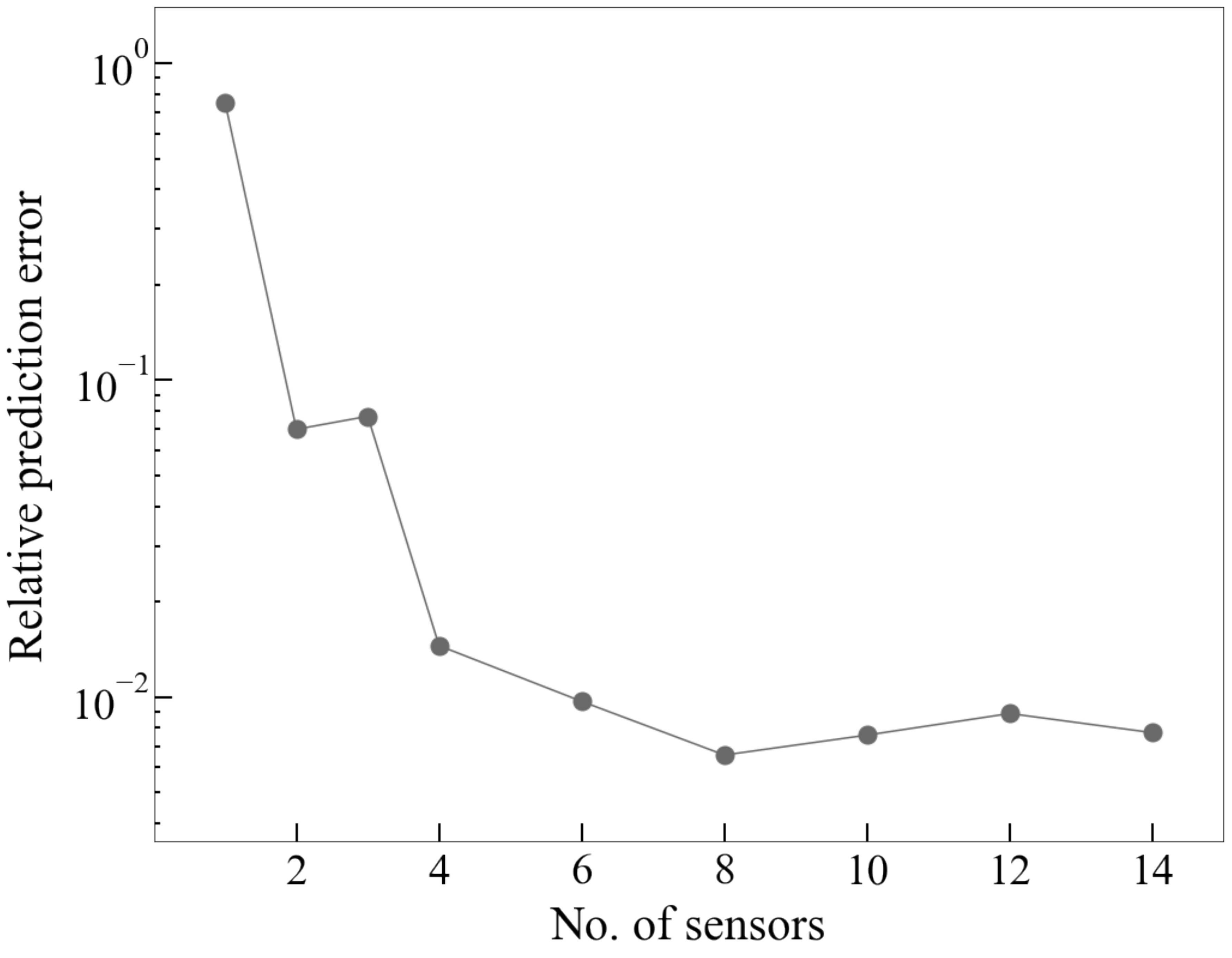}
    \centering
    \caption{{\bf Relative error of magnetic field} Relative prediction error of magnetic field versus different numbers of sensors, with the vertical axis in logarithmic scale. The error plummets when the number of sensors increases from 1 to 6. Afterward, the error stays in the same order of magnitude and varies non-systematically.}
    \label{Fig3}
\end{figure}

To have a better understanding of the relation between Maxwell's equations and the NNs, we look at how the error depends on the number of sensors. As shown in Fig.~\ref{Fig3}, the error drastically reduces when the number of sensors increases to 6 and has a similar value onward. This dramatic reduction can be explained by the number of boundary conditions required to identify the hyperplane from directly calculating Maxwell's equations. $\nabla^{2}\mathbf{B} - \frac{1}{c^{2}}\frac{\partial^2 \mathbf{B}}{\partial t^{2}} = -\mu \nabla \times \mathbf{J}
    \label{Max}$  shows the form of the time-dependent Maxwell's equations that only depend on the magnetic field. Through this, we can count the number of integration constants and thus, the boundary conditions required to get a unique solution by multiplying the number of independent variables, the order of differential equation, and the number of components together. For the Maxwell's equation, the obtained value is 24, but we are only concerning time-independent cases, therefore resulting required number of 18. Since the boundary conditions are provided by magnetic field measured at different spatial points by sensors, the minimum number of sensors, each has three components, needed to obtain a unique and accurate result is equal to 6. This, together with the claim that the NN is approximating the ideal hyperplane calculated from Maxwell's equations, explains the rapid reduction of error in Fig.~\ref{Fig3}. When below 6 sensors are used, the NN does not have enough information needed for the ideal hyperplane to calculate the unique result. Therefore, the trend in Fig.~\ref{Fig3} indicates a strong relation between the NN and Maxwell's equations, strengthening the claim that training the NN is equivalent to bringing the NN closer to the ideal hyperplane calculated from Maxwell's equations. However, it is crucial to notice the NN can only approximate the ideal hyperplane but never reach it since their mathematical forms are fundamentally different. Hence, errors always exist from the network's prediction. Besides, the quality of the approximation greatly depends on and is limited by the training conditions and data. This aspect of the network will be clearly shown in the following result.

\section*{NN's performance over a range of magnetic field}
In this study, we evaluate the performance of the neural network under various magnetic fields created by the coils $B_{control}$, ranging from 0.6~G to 100~G. These values cover the typical range of magnetic fields used in experiments involving ultracold dipolar atoms~\cite{seo2023apparatus}. This can be achieved by generating multiple prediction data sets with different average magnitudes of output. Before commencing the test, it is crucial to consider the average magnitude of the output of the training data since it could significantly affect the training process that determines the network's response to different magnetic fields.

\begin{figure*}
\centering
\includegraphics[width=1\linewidth]{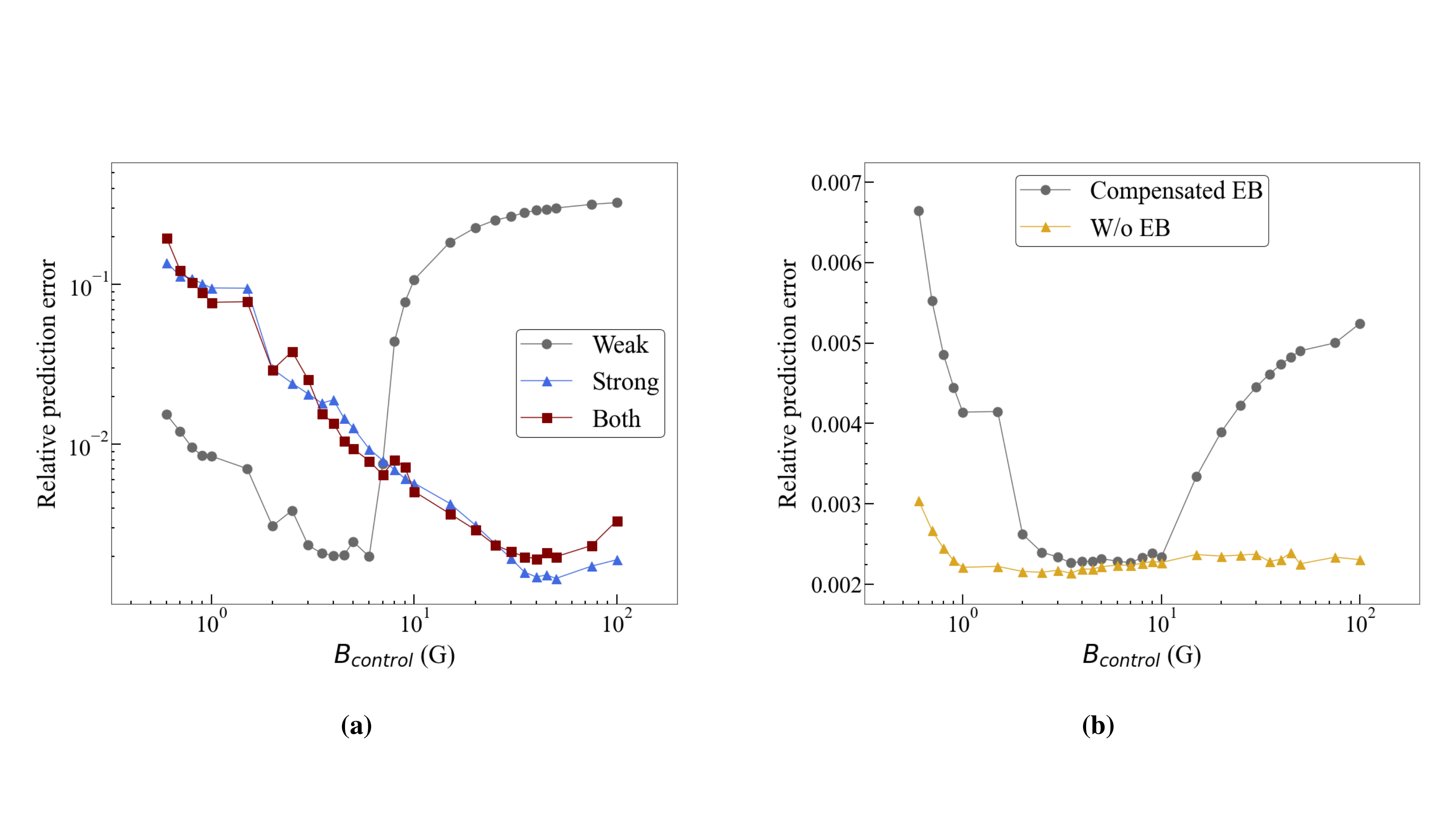}
\caption{{\bf Relative prediction error for various conditions} Relative prediction error versus the magnetic field generated by the coils $B_{control}$, covering from 0.6 to 100~G range. (a) The Earth's magnetic field $B_{Earth}$ is present and compensated (assumed to be 400~mG). The legend shows in which range of field the artificial NN is trained. Weak refers to a field from 1 to 5~G and strong refers to 10 to 50~G range. The error is minimized and reduced to around 0.2$\%$ over a range where the training field has a similar magnitude as the testing field. However, the error increases rapidly to over 10$\%$ outside the working range. The working range of the NN trained with a stronger field is wider (in log scale) than the one trained with a weak field. When the NN is trained with both weak and strong fields, the result is similar to the one produced by a NN trained with a strong field only. (b) $B_{Earth}$ (EB in the figure) is either compensated or removed entirely as indicated by the legend. For the one with compensated $B_{Earth}$, the NN is trained with a weak field. The minimized region still remains but the increase in error is way less drastic outside the range compared to having full $B_{Earth}$. The error can be suppressed under 0.7$\%$ over the whole range. For the one without $B_{Earth}$, the NN is trained at 2.5~G. The performance of the network is very consistent over the whole range with errors around or below 0.3$\%$.}
    \label{Fig4}
\end{figure*}

Fig.~\ref{Fig4}(a) demonstrates the NN's accuracy in predicting the magnetic field at the center of the chamber under different magnetic field conditions.
When the NN is trained with weak magnetic field strength ($B_{control}$) of 1-5 G, the error is minimized within that range and can be reduced to as low as approximately 0.2$\%$. However, when the NN is trained with a strong magnetic field strength of 10-50 G, the error is minimized at a $B_{control}$ order of magnitude of 10 G. In addition, when the $B_{control}$ is outside of the specified range, the error quickly surges to above 10$\%$. This issue could be avoided if the magnetic field were completely rescalable. Unfortunately, this is not possible due to the presence of the constant Earth magnetic field ($B_{Earth}$). As $B_{control}$ changes, so does $B_{Earth}/B_{control}$, and these variations can be drastic, particularly when $B_{control}$ changes by an entire order of magnitude. If the NN is trained with a weak magnetic field strength, there is a possibility that the network would calculate the bias based on $B_{Earth}/B_{control}$ for a weaker field even when $B_{control}$ is much stronger, leading to a higher error. This also explains why the applicable range of $B_{control}$ for NN trained at weaker $B_{control}$ is narrower, as $B_{Earth}/B_{control}$ varies much faster at weaker magnetic field strengths.

To expand the applicable range of the neural network, one option is to train it on a larger range of $B_{control}$. However, this method has been proven to be ineffective, as shown in Fig.~4(a). The figure demonstrates that the errors between the neural network trained with a large range of $B_{control}$ and the one trained with strong $B_{control}$ are very similar. This indicates that the neural network simply ignores weaker $B_{control}$. The reason for this lies in the use of the RMSE loss function in absolute value. Larger $B_{control}$ usually produces larger absolute errors, causing the network's parameters to be tuned in a way that provides more accurate results at larger $B_{control}$ over weaker ones.

Here we suggest two viable ways to improve the working range of the NN. First, the NN can simply be trained without $B_{Earth}$ in the data by adjusting the sensors offsets until the reading is zero or compensating $B_{Earth}$ in the laboratory. Both of the methods involve removing the effect of the $B_{Earth}$ during the training process. In Fig.~\ref{Fig4}(b), the error of the network's prediction can then be kept below 0.4$\%$ over the whole range when there is no $B_{Earth}$. Therefore, this method is very effective in principle but requires good control over the sensors.

The second method is to compensate $B_{Earth}$ using coils in the laboratory to attenuate the effect of $B_{Earth}$. Practically, it is difficult to completely eliminate $B_{Earth}$, but it is possible to compensate 90$\%$. Therefore, we evaluate the effectiveness of this method by training NN with a small bias of 40~mG, which is displayed in Fig.~\ref{Fig4}(b). The error is below 0.7$\%$ over the whole range and below 0.3$\%$ from 20~G to 100~G, indicating a good precision and wide working range can be achieved simultaneously with this method.

Nonetheless, even when both of the above methods are not applicable, the NN is still useful and powerful. As long as the approximate value of the magnetic field is known, the properly trained NN can always be picked to calculate the magnetic field with an extremely low error. When the order of the magnetic field is unknown, however, we can still use the NN trained with the wrong range to find back the order of magnitude since the error, in this case, is still over an order larger. The more accurate result can then be calculated using the properly trained NN. 

\section*{Application to cold atoms experiments}
The performance of magnetic field regression demonstrated in this work opens a new possibility of applying this method to experiments with cold atoms. One such system is an apparatus of dipolar erbium atoms which have a large magnetic moment of 7$\mu_B$ where $\mu_B$ is the Bohr magneton and contains a dense spectrum of Feshbach resonances~\cite{frisch2014quantum}. Even at a low magnetic field regime, erbium atoms have multiple Feshbach resonances for a magnetic field smaller than 3~G, where the widths of these resonances are around 10-200~mG~\cite{aikawa2012bose}. The regression method could allow us to monitor the magnetic field vector with the resolution of $\sim$10~mG in the range of 3~G, which is accurate enough to properly calibrate the experimental system. This is still favorable for other atomic species, such as alkali atoms, since a maximal error of magnetic field is about 1.25~G in the range of 500~G, which is sensitive enough compared to the linewidth of most commonly used Feshbach resonances~\cite{chin2010feshbach}. However, even though the regression method can provide useful results, we suggest that its accuracy should be verified with another method, such as radio-frequency spectroscopy.

Apart from precisely determining the magnetic field inside a vacuum chamber, the proposed method also serves as a quick indicator for magnetic field change. For instance, when sensors around the vacuum chamber give unexpected values, it indicates a change in a magnetic field. If all 6 sensors are changed along the same direction, then an external magnetic field is suddenly introduced to the system. On the other hand, if these 6 sensors change in a different direction, then it is likely because the position of a device within the area is changed. The former can be solved by compensating the external magnetic field by coils, while the latter requires to re-train the NN. The change of sensor value signals a change in the environmental magnetic field, which provides important information and is easy to monitor.

\section*{Conclusion}
In conclusion, a novel method based on NN to precisely determine magnetic fields in an isolated vacuum environment is demonstrated. An artificial NN is trained such that it outputs the magnetic field at the center of the vacuum chamber based on the magnetic fields surrounding the chamber. The effect of the number of sensors and the magnitude of the magnetic field on the performance of the trained NN are evaluated.

After training, the relative error of magnetic field magnitude below 0.3$\%$ is achieved under a wide range of magnetic fields, which is sufficient to calibrate the magnetic field in our erbium quantum gas apparatus, where many narrow Feshbach resonances exist even in the low field regime. Besides, experiments with other atomic species can benefit from this method. Furthermore, as no special setup is required, the established method can be extended to other magnetic field-sensitive experiments conducted in an isolated environment.

\vspace{5pt}
\paragraph*{Acknowledgments}
We acknowledge support from the RGC through 16306119, 16302420, 16302821, 16306321, 16306922, C6009-20G, N-HKUST636-22, and RFS2122-6S04. 

\vspace{5pt}


\end{document}